\documentclass[12pt]{iopart}
\usepackage {graphicx}

\begin{document}
\title[J. Vahedi et all. ]{Entanglement study of the 1D Ising model with Added Dzyaloshinsky-Moriya interaction}
\author{J. Vahedi$^{1}$, M. R. Soltani$^{2}$ and S. Mahdavifar$^{3}$}
\address{ $^{1}$ Department of Physics, Science and Research Branch of Islamic Azad University, Tehran, Iran.\\
$^{2}$ Department of Physics, Shah-ray Branch of Islamic Azad University, Tehran, Iran.\\
$^{3}$ Department of Physics, University of Guilan, 41335-1914, Rasht, Iran.}
\ead{javahedi@iausari.ac.ir,\quad smahdavifar@gmail.com}
\date{\today}

\begin{abstract}

We have studied occurrence of quantum phase transition in the one-dimensional spin-1/2 Ising model with added Dzyaloshinsky-Moriya (DM) interaction
from bi-partite and multi-partite entanglement point of view. Using exact numerical solutions, we are able to study such systems up to $24$ qubits.
The minimum of the entanglement ratio $R\equiv\tau_{2}/\tau_{1}<1$, as a novel estimator of QPT, has been used to detect QPT  and our calculations
have shown that its minimum took place at the critical point. We have also shown both the global-entanglement (GE) and multipartite entanglement (ME) are maximal at the critical point for the Ising chain with added DM  interaction.
Using matrix product state approach, we have calculated the tangle and concurrence of the model and it is able to
capture and confirm our numerical experiment result. Lack of inversion symmetry in the presence of DM interaction stimulated
us to study entanglement of three qubits in symmetric and antisymmetric way which brings some surprising results.
\end{abstract}
\pacs{75.10.Jm; 75.10.Pq}
\submitto{\JPCM}

\maketitle

%########################################################################
%######################                        ##########################
%######################      Section I     ##########################
%######################                        #########################
%#######################################################################

\section{INTRODUCTION}\label{sec1}

In the last few years it has become apparent that quantum
information may lead to further insight into other
areas of physics such as condensed
matter and statistical mechanics \cite{a1,a2,a3,a4,a5,a6,a7,a8,a9,b1,b2,b3}.
The attention of the quantum information community to study in condensed
matter has stimulated an exciting cross fertilization
between the two areas. It has been found that entanglement
plays a crucial role in the low-temperature physics of
many of these systems, particularly in their ground
state\cite{a4,a5,a6,a7,b1}.
Quantum phase transition (QPT) happens  at zero temperature
and shown  non analyticity in
the physical properties of the ground state by the change of a
parameter $\lambda$ of the
Hamiltonian $H(\lambda)$. This change is driven
only by quantum fluctuations\cite{Sachdev}.
Since QPT occurs at $T=0$, the emerging correlations have
a purely quantum origin. Therefore, it is reasonable to
conjecture that entanglement is a crucial ingredient for
the occurrence of the QPT \cite{a4,a6,a7,c2,Wooters98,c8}.
%If this is true, then the QPT would imprint its
%signature on the behavior of an entanglement measure.
Wu et al. \cite{a7} have shown that a discontinuity in a bipartite
entanglement measure (concurrence\cite{Wooters98} and negativity \cite{c8})
is a necessary and sufficient indicator of a first-order
quantum phase transition, negativity being characterized
by a discontinuity in the first derivative of the
ground state energy. They have also shown that a
discontinuity or a divergence in the first derivative of the
same measure (assuming it is continuous) is a necessary
and sufficient indicator of a second-order QPT,
that is characterized by a discontinuity or a divergence of
the second derivative of the ground state energy.
\par
Dzyaloshinsky has shwon\cite{Dzyaloshinsky58}  that, in crystal with
no inversion center, the usual isotropic exchange $J\vec{S}_{i}.\vec{S}_{j}$
is not the only magnetic interaction and antisymmetric exchange
$\vec{D}_{ij}.(\vec{S}_{i}\times\vec{S}_{j})$
is allowed. Later, Moriya has shown\cite{Moriya60}that
inclusion of spin orbit coupling on magnetic ions in 1st and 2nd order
leads to antisymmetric and anisotropic exchange respectively.
This interaction is, however, rather difficult to handle analytically,
but it is one of the agents responsible for magnetic frustration. Since this interaction may
induce spiral spin arrangements in the ground state\cite{sa6}, it
is closely involved with ferroelectricity in multiferroic spin
chains\cite{sa7,sa8}. Besides, the DM interaction plays an important
role in explaining the electron spin resonance experiments
in some one-dimensional antiferromagnets\cite{sa9}. Moreover,
the DM interaction modifies the dynamic properties\cite{sa10}
and quantum entanglement\cite{sa11} of spin chains\cite{sa12}.
 In the present paper, we are interested to study the one-dimensional spin-1/2
 Ising model with added DM interaction from quantum entanglement point of view
 using variational matrix product state and
 numerical exact diagonalization methods. The Hamiltonian is given by
 \begin{eqnarray}
\emph{H}=J\sum_{j=1}^{N}S_{j}^{z}S_{j+1}^{z}+\sum_{j=1}^{N} \vec{D}.\left(\vec{S}_{j}\times \vec{S}_{j+1}\right),
\label{ISING}
\end{eqnarray}
where $\overrightarrow{S}_{j}$ is spin-$1/2$ operator on the $j$-th site, and $J>0$ $(J<0)$ denotes  antiferromagnetic (ferromagnetic)
coupling constant. In very recent works\cite{soltani09, Jafari08}
respectively studied the ground state phase diagram of the ferromagnetic and antiferromagnetic Ising chain in
the presence of the uniform DM interaction. It is found that the ground state phase diagram of both
systems consists of spiral-ferromagnetic and spiral-antiferromagnetic phases respectively and a commensurate-incommensurate
(C-IC) quantum phase transition occurs at $D_c=|J|$. However at the critical value $D_c$, a
metamagnetic phase transition occurs into the chiral gapless phase in the ground state phase diagram of the ferromagnetic chain.

This paper is structured as follows. In section II,
we will discuss about bipartite and multipartite
entanglement measures as QPT indicators of our model
and we will present our numerical study. In section III, the
variational matrix product states will be outlined and bipartite
entanglement will be obtained. In section IV, we willstudy entanglement of
three qubits in symmetric and antisymmetric way.
Finally we conclude and summarize our results in section V.
%########################################################################
%######################                        ##########################
%######################      Section II       ##########################
%######################                        #########################
%#######################################################################
\section{QUANTUM PHASE TRANSITION}\label{sec2}

\subsection{Bipartite Entanglement}\label{subec1}

The occurrence of collective behavior in many-body quantum systems is associated with
classical and quantum correlation. The quantum correlation, which known as entanglement, cannot be
measured in terms of classical physics and represents the impossibility of giving a
local description of many-body quantum state.
The issue of finding entanglement measures has recently attracted an increasing
interest\cite{a5,a6,TRos04,TRos05}. Concurrence and tangle are the most widely
used measures in QPT related entanglement studies.
Both of these measures are for bipartite states and because of monogamous nature
of entanglement they are expected to decrease at the quantum critical point, if entanglement is shared by the whole system.
Therefore, in order to manifest the presence of (QPT) in the model described by Eq.(\ref{ISING}),
 we focus on the entanglement of formation \cite{Benn96} in the quantum spin system
 and make use of the  \emph{one-tangle} and of the concurrence.
 The one-tangle \cite{Amico04,VCoffman00} quantifies the zero temperature
 entanglement of a single spin with the rest of the system and defines as
\begin{eqnarray}
\tau_{1}=4\det\rho^{(1)},\quad \rho^{(1)}=\frac{1}{2}(I+\sum_{\alpha}M^{\alpha}S^{\alpha}),
\label{tangle}
\end{eqnarray}
where $\rho^{(1)}$ is the one-site reduced density matrix,
$M^{\alpha}=\langle S^{\alpha}\rangle$, and $\alpha=x, y, z$. In terms of the spin
expectation values $M^{\alpha}$, one has:
\begin{equation}
\tau_{1}=1-4\sum_{\alpha}(M^{\alpha})^{2}.
\label{tangle2}
\end{equation}
On other hand, the concurrence\cite{Wooters98} quantifies
instead the pairwise entanglement between two spins and defines as
%***********************************************************
\begin{eqnarray}
C_{lm}&=&2~max\{0, C_{lm}^{(1)}, C_{lm}^{(2)}\},
\label{concurrence}
\end{eqnarray}
%***********************************************************
where
%***********************************************************
\begin{eqnarray}
C_{lm}^{(1)}&=& \sqrt{(g_{lm}^{xx}-g_{lm}^{yy})^{2}+(g_{lm}^{xy}+g_{lm}^{yx})^{2}} \nonumber \\
&-&\sqrt{(\frac{1}{4}-g_{lm}^{zz})^{2}-(\frac{M_{l}^{z}-M_{m}^{z}}{2})^{2}}\nonumber \\
C_{lm}^{(2)}&=& \sqrt{(g_{lm}^{xx}+g_{lm}^{yy})^{2}+(g_{lm}^{xy}-g_{lm}^{yx})^{2}} \nonumber \\
&-&\sqrt{(\frac{1}{4}+g_{lm}^{zz})^{2}-(\frac{M_{l}^{z}+M_{m}^{z}}{2})^{2}}
\label{concurrence}
\end{eqnarray}
%***********************************************************
and $g_{lm}^{\alpha\beta}=\langle S_{l}^{\alpha}
S_{m}^{\beta}\rangle$ is the correlation function between spins on
sites $l$ and $m$ and $M_{l}^{z}=\langle S_{l}^{z} \rangle$.
The notation $\langle ... \rangle$ represents the ground
state expectation value.
%%%%%%%%%%%%%%%%%%%%%%%%%%%%%%%%%%%%%%%%%%%%%%%%%
%%%%%%%%%%   FIG-1   %%%%%%%%%%%%%%%%%%%%%%%%%%
\begin{figure}
  \begin{center}
   \includegraphics[height=80mm,width=100mm]{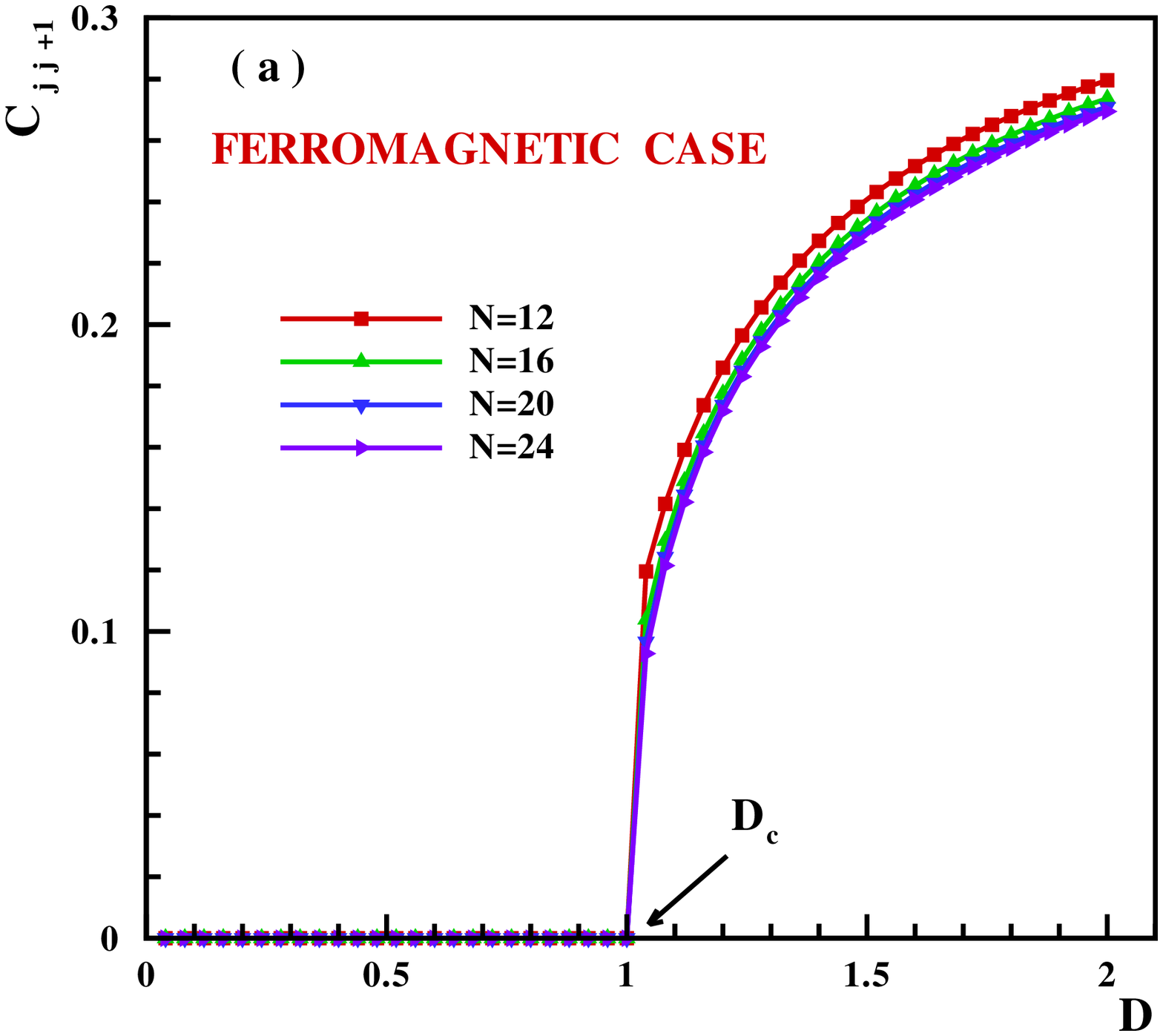}\\
    \includegraphics[height=80mm,width=100mm]{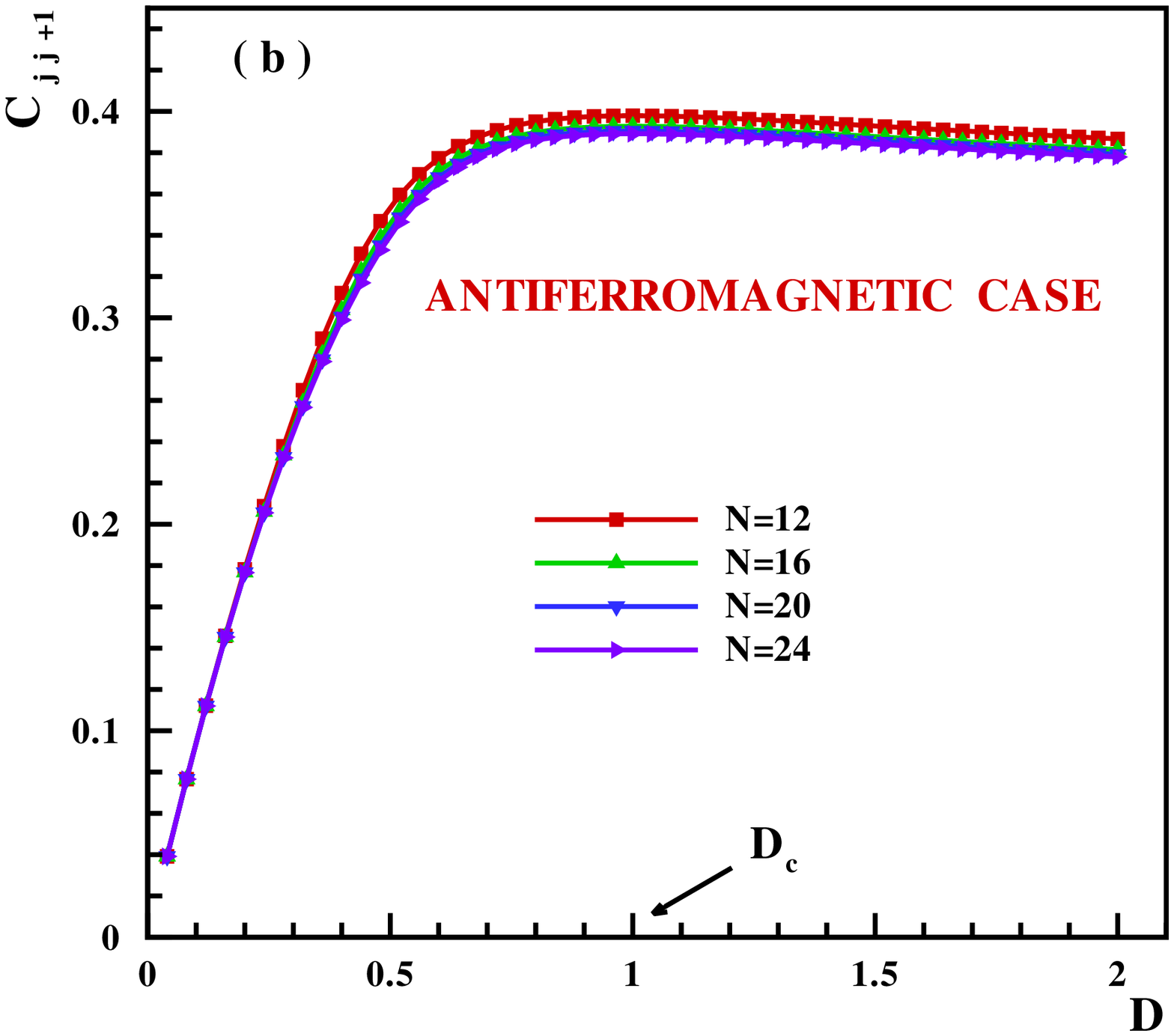}\\
  \caption{(Color online.) The concurrence is plotted as a function
of DM vector $D$, (a) ferromagnetic and (b) antiferromagnetic cases for
different lengths N=12, 16, 20, 24.}\label{C-F}
\end{center}
\end{figure}

%%%%%%%%%%%%%%%%%%%%%%%%%%%%%%%%%%%%%%%%%%%%%%%
%%%%%%%%%%%%%%%%%%%%%%%%%%%%%%%%%%%%%%%%%%%%%%%
One-tangle and concurrence are related by Coffman-Kundu-Wootters (CKW) conjecture \cite{VCoffman00},
which had been proved by Osborne and Verstraete \cite{osborne06}, stating that
\begin{equation}
\tau_{1}\geq\tau_{2}\equiv\sum_{l\neq m}C_{lm}^{2}.
\end{equation}
Which expresses the crucial fact that pairwise entanglement does not exhaust the global entanglement
of the system, as entanglement can also be stored in 3-spin correlations, 4-spin correlations, and so on.
Authors in Ref.(\cite{TRos04,TRos05}), have proposed that, due to CKW conjecture, the minimum of the entanglement ratio
$R\equiv\tau_{2}/\tau_{1}<1$, as a novel estimator of QPT, fully based on entanglement quantifiers.

By doing an experiment, one can find a clear picture of the entanglement
phenomenon in the ground state magnetic phases of the model. Since a real experiment
cannot be done at zero
temperature, the best way is doing a virtual numerical experiment.
A very famous and accurate method in the field of the numerical experiments is
known as the Lanczos method. However, the strong role of a numerical
experiment to examine quantum phase transitions is not negligible.
To explore the nature of the entanglement in different magnetic phases,
we used Lanczos method to diagonalize
numerically chains with length up to $N=24$ and coupling constant $|J|=1$.
The ground state eigenvector, $|Gs\rangle$,
was obtained for chains with periodic
boundary conditions. The numerical Lanczos results on the concurrence for
the Ising chain with
DM interaction, are shown in Fig.~\ref{C-F}. As is clearly seen
from Fig.~\ref{C-F}(a), in the case of ferromagnetic chain and for $D<D_c=|J|=1$,
the concurrence is equal to zero which shows that the ground state of
the system is in the fully non-entangled polarized ferromagnetic phase. At the critical
value $D_c=|J|=1$, the concurrence jumps to a non-zero value which confirms
the metamagnetic phase transition. By more increasing the DM vector, $D>D_c$,
the ground state is in the chiral phase and nearest neighbors are entangled.
On the other hand, in the case of the antiferromagnetic Ising model, as can be clearly seen
from Fig.~\ref{C-F}(b), in the absence of the DM interaction, the ground state
is non-entangled which is related to the saturated N$\acute{e}$el phase. As
soon as the DM vector applies, nearest neighbors will be entangled
and concurrence between them increases from zero. Thus in the case of
antiferromagnetic Ising chains, the DM interaction induces the quantum
correlations of the two spins and nearest neighbor spins will be entangled
as soon as the DM interaction applies. In contrast, in the case of the
ferromagnetic Ising chains the DM interaction only induces the quantum
correlations after the critical value $D_c$ and nearest neighbor spins
will not be entangled up to the critical value $D_c$.

\begin{figure}
  \begin{center}
   \includegraphics[height=80mm,width=100mm]{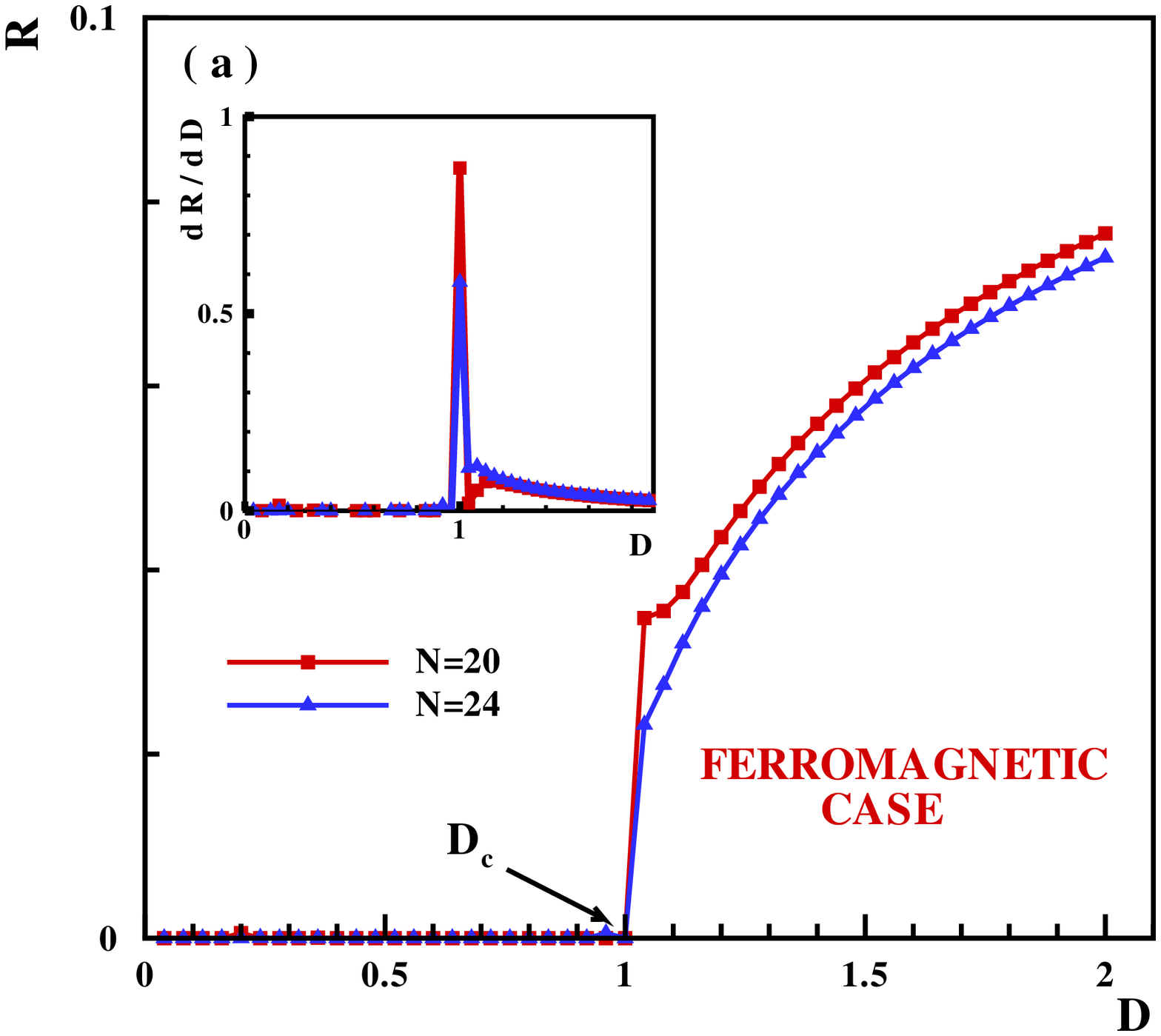}\\
    \includegraphics[height=80mm,width=100mm]{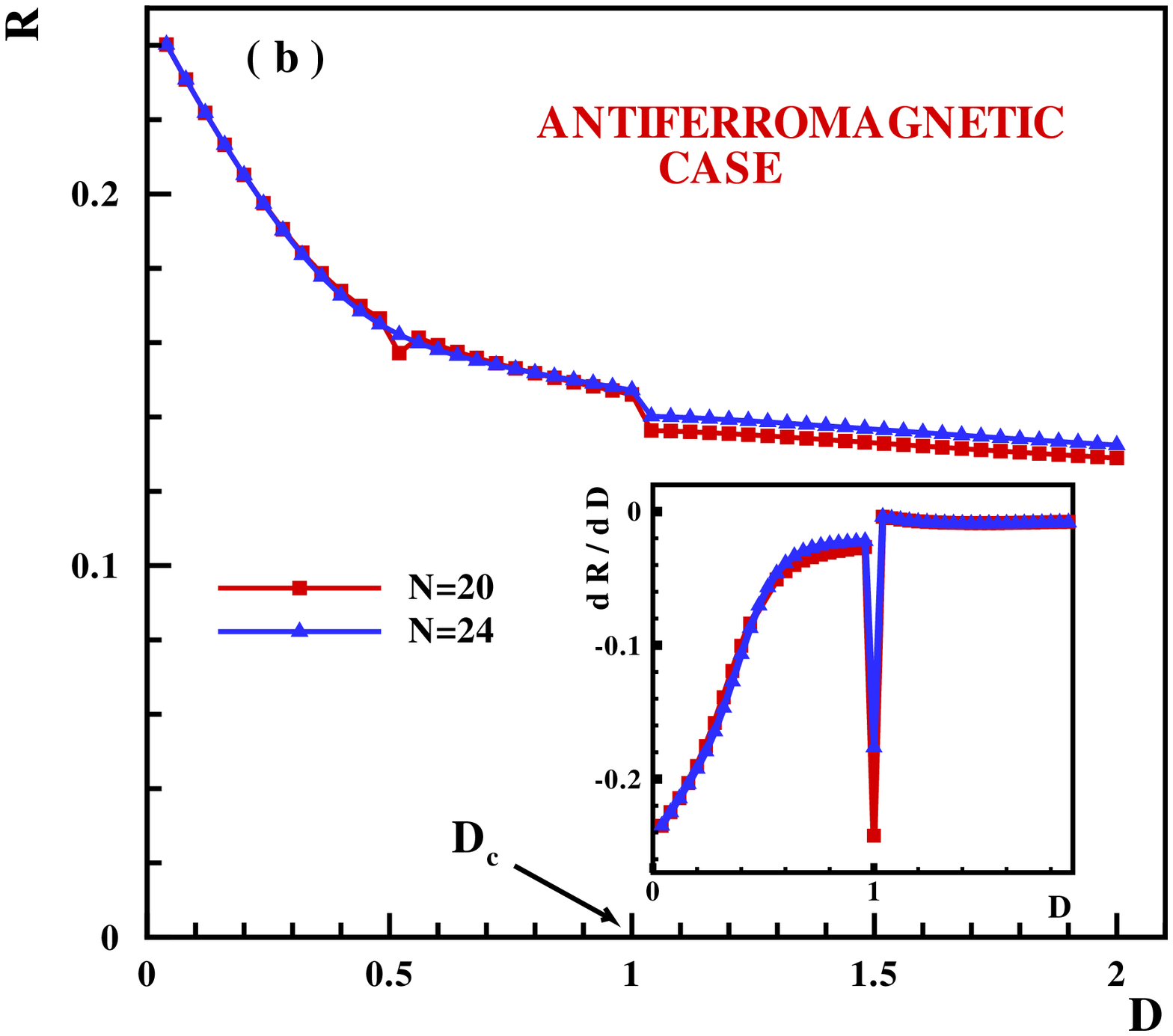}\\
  \caption{ (Color online.) The entanglement ratio
 $\tau_{1}/\tau_{2}=R$ is plotted as a function
 of DM vector $D$, (a) ferromagnetic and (b) antiferromagnetic cases for
different lengths N=20, 24. Inset: the first derivative
 of entanglement ratio. }\label{C1-F}
\end{center}
\end{figure}

An additional insight into the nature of different phases can be
obtained by studying the entanglement ratio. Therefore, we have calculated
the entanglement ratio by using Lanczos method in both ferromagnetic and antiferromagnetic cases.
We have plotted our numerical results
in Fig.~\ref{C1-F}. For ferromagnetic case, as it can be seen from Fig.~\ref{C1-F}(a),
the entanglement ratio remains zero up to the critical DM interaction $D_c$ which
is expected from the saturated ferromagnetic phase. As soon as the DM interaction
increases from the critical $D_c$, the entanglement ratio starts to increase from zero.
In the inset of Fig.~\ref{C1-F}(a), the first derivative entanglement ratio is plotted.
As it is seen in the ferromagnetic phase, $D<D_c$, the derivative is equal to zero and an abrupt change
took place exactly at $D_{c}=|J|=1$ which is an indication
of the quantum phase transition.
In the antiferromagnetic case,  Fig.~\ref{C1-F}(b), as soon as the DM interaction applies the
entanglement ratio creates and decreases by increasing the DM vector up to $D_{c}=1.0$
which a change took place and in the $D>D_{c}$ region the ratio becames monotonous.
 In the inset of Fig.~\ref{C1-F}(b), the first derivative entanglement ratio is plotted.
 It is clear that in the $D>D_{c}$ region derivative of ratio is equal to zero
and an abrupt change took place exactly at $D_{c}=|J|=1.0$ which is an indication
of the quantum phase transition. 

%The oscillations of ratio at finite N in the region
%$D<D_{c}$ is the result of level crossing between the ground state and excited states of
%the model.

%%%%%%%%%%%%%%%%%%%%%%%%%%%%%%%%%%%%%%%%%%%%%%%%%
%%%%%%%%%%   FIG-2   %%%%%%%%%%%%%%%%%%%%%%%%%%

%%%%%%%%%%%%%%%%%%%%%%%%%%%%%%%%%%%%%%%%%%%%%%%%%
%%%%%%%%%% %%%%%%%%%%%%%%%%%%%%%%%%%%%%%%%%%%%%%

\subsection{Global Entanglement}\label{subec2}

Because of many different kinds of entanglement, quantifying of multipartite entanglement states
(MES) is more difficult. Global-entanglement ($E_{gl}$) measure defined by Meyer and Wallach \cite{MW}
which can measure the total nonlocal information per particle in a general multipartite system\cite{Afshin}
\begin{equation}
E_{gl}=\frac{1}{N}\left[2\sum_{i_{1}<i_{2}}\tau_{i_{1}i_{2}}+\cdot\cdot\cdot+N\sum_{i_{1}<\cdot\cdot\cdot<i_{N}}\tau_{i_{1}\cdot\cdot\cdot i_{N}}\right].
\end{equation}
$E_{gl}$ is the average of tangles per particles ($\frac{<\tau>}{N}$ ), without
giving detailed knowledge of tangle distribution among the
individual particles. Therefore, $E_{gl}$ is an average quantity
and cannot distinguish between entangled states which have equal $ <\tau>$ yet different
distributions of tangles, like $GHZ_{N}$ (Greenberger-Horne-Zeilinger)
 and $EPR^{\otimes N/2}$ state. $E_{gl}$  has ability to discriminate
 between $GHZ$ from $W$ states because of their different values of tangle.
De Oliveira et al. \cite{rigolin} also
introduced a slight extension of global entanglement as generalized global entanglement (GGE)
which, in contrast to global entanglement, the proposed GGE measure can distinguish three paradigmatic entangled
$GHZ_{N}$ (Greenberger-Horne-Zeilinger), $EPR^{\otimes N/2}$ and W states.
\begin{eqnarray}
G(2,n)&=&\frac{4}{3}\frac{1}{N-1}\sum^{N-1}_{n=1}\left[1-\frac{1}{N-1}\sum^{N}_{j}Tr\rho^{2}_{j,j+n}\right]\nonumber\\
      &=&\frac{4}{3}\left[1-\frac{1}{4}\sum_{\alpha,\beta=0}^{3}\big<\sigma_{j}^{\alpha}\sigma_{j+1}^{\beta}\big>^{2}\right].
      \label{GGE}
\end{eqnarray}

\begin{figure}
  \begin{center}
   \includegraphics[height=80mm,width=100mm]{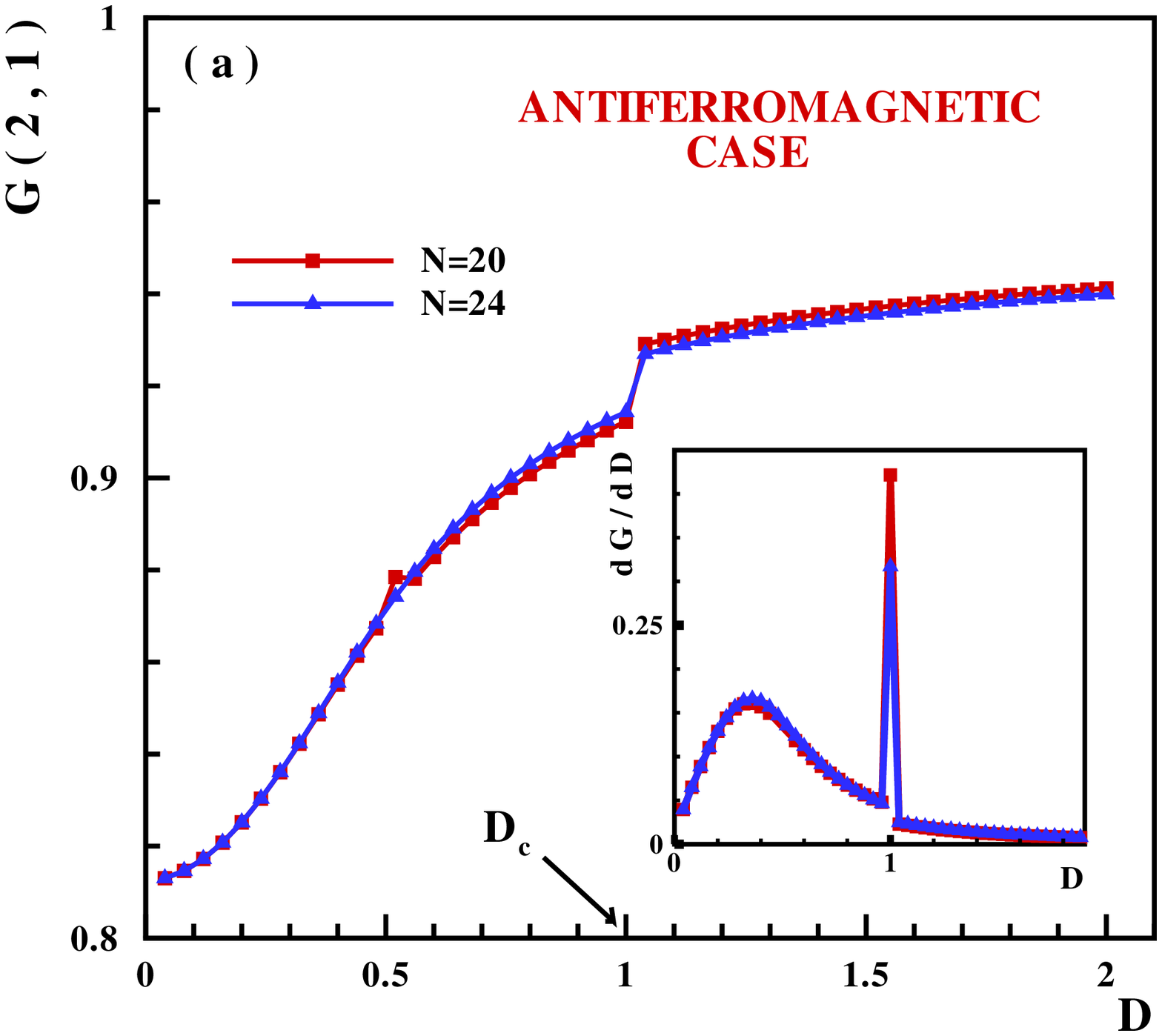}\\
    \includegraphics[height=80mm,width=100mm]{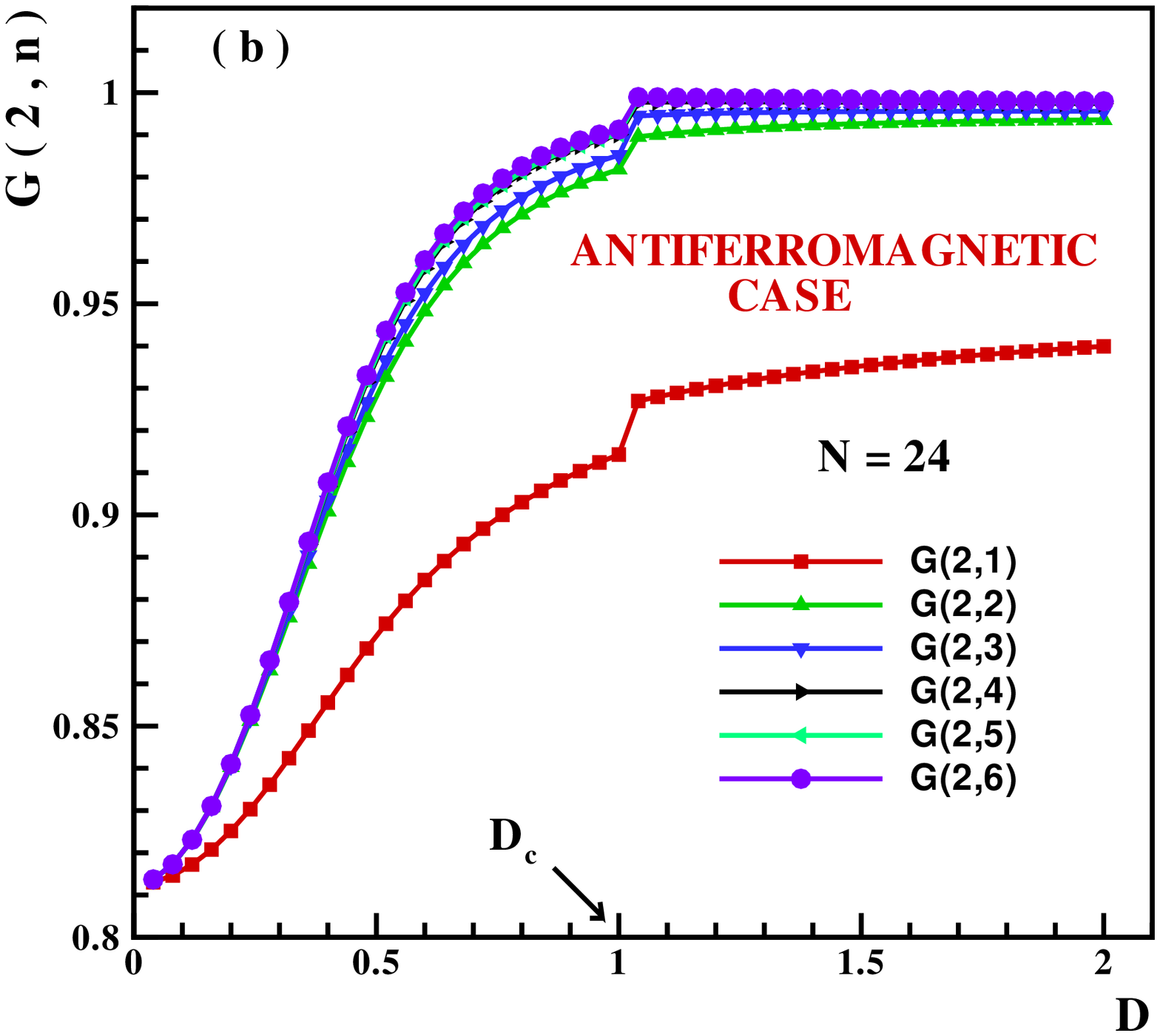}\\
  \caption{(Color online.) (a) Multipartite entanglement estimator
$G(2,n)$ is plotted as a function of DM vector $D$ for antiferromagnetic Ising chain with length N=20, 24. Inset:
the first derivative of multipartite entanglement$\frac{dG(2,n)}{dD}$. (b)
$G(2,n)$ for the chain size N=24, is plotted in the limite of $n\rightarrow\infty$.}\label{multipartite}
\end{center}
\end{figure}

\begin{figure}
  \begin{center}
   \includegraphics[height=80mm,width=100mm]{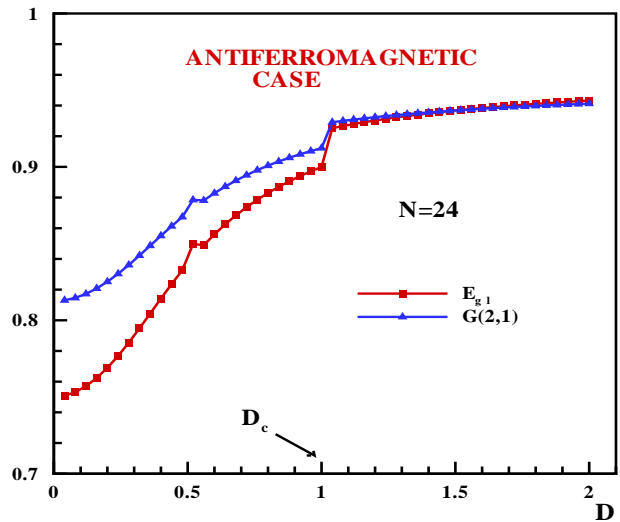}\\
    \caption{(Color online.) Multipartite entanglement $G(2,n)$
and global entanglement $E_{gl}$ are plotted as a function
of DM vector $D$, for antiferromagnetic Ising chain with length N=24.}\label{multipartite3}
\end{center}
\end{figure}

As such the generalized measure can detect a genuine multipartite entanglement and is maximal at the critical point\cite{rigolin}.
Here, we have calculated multipartite entanglement Eq.(\ref{GGE}) by using Lanczos numerical method with periodic boundary
conditions in antiferromagnetic case. We have plotted our numerical results
in Fig.\ref{multipartite} for antiferromagnetic case. As it can be seen from Fig.\ref{multipartite}(a) the
multipartite entanglement starts to increase by increasing
DM  up to the critical DM interaction $D_c$ which a change took place exactly
at $D_{c}=|J|=1$ and then after that the multipartite entanglement reaches the saturation value.
Our calculation shows that multipartite entanglement is maximal around the critical point
$D_{c}$. In order to get better insight into the ability of  multipartite entanglement as
a quantum phase transition toolkit we have plotted the first derivative entanglement
multipartite entanglement in the inset of Fig.~\ref{multipartite}(a). It shows
divergent behavior at the critical point $D_{c}$.
We have also plotted $lim_{n\rightarrow\infty} G(2,n)$ in Fig.~\ref{multipartite}(b).
Our calculation shows $G(2,n)$ increases as $n\rightarrow\infty$ at the critical point.
Here, we have finally calculated global entanglement $E_{gl}$ of model Eq.(\ref{ISING}) and
compar it with multipartite entanglement.
It can be seen from Fig.~\ref{multipartite3} that both $G(2,n)$ and $E_{gl}$ are maximal
at the critical point $D_c$ and their behavior is qualitatively the same.

%%%%%%%%%%%%%%%%%%%%%%%%%%%%%%%%%%%%%%%%%%%%%%%%%
%%%%%%%%%%   FIG-3   %%%%%%%%%%%%%%%%%%%%%%%%%%

%%%%%%%%%%%%%%%%%%%%%%%%%%%%%%%%%%%%%%%%%%%%%%%%%
%%%%%%%%%%   FIG-4   %%%%%%%%%%%%%%%%%%%%%%%%%%

%########################################################################
%######################                        ##########################
%######################      Section III       ##########################
%######################                        #########################
%#######################################################################

\section{VARIATIONAL MATRIX PRODUCT STATE APPROACH} \label{sec3}

The matrix product state is defined as\cite{TH91, MF89}
\begin{eqnarray}
 |\psi\rangle=Tr(g_{1},g_{2},...,g_{N}),
\end{eqnarray}
 where $g_{j}=a_{j}|\uparrow\rangle_{j}+b_{j}|\downarrow\rangle_{j}$, and $a_{j}$ and $b_{j}$
 are probability amplitudes for two spin configurations at site $j$.  In what follow, we intend  to determine the ground state energy of ferromagnetic
  Ising spin system with DM interaction. In this respect, by using the above formalism,
 the variational energy is obtained by
 \begin{eqnarray}
E_{var}=\langle H \rangle=\frac{\langle\psi|H|\psi\rangle}{\langle \psi|\psi\rangle}=\sum_{j}\frac{\hat{H}_{j,j+1}}{[G_{j},G_{j+1}]},
\label{var}
\end{eqnarray}
where $\langle \psi|\psi\rangle=g_{1}\otimes g_{1}...g_{N}\otimes g_{N}=\Pi_{j}G_{j}$ and
$G_{j}=g_{j}\otimes g_{j}=|a_{j}|^{2}+|b_{j}|^{2}$ here $\hat{H}_{jk}=J\hat{S}_{j}^{z}\hat{S}_{k}^{z}+\vec{D}.(\vec{S}_{j}\times\vec{S}_{k})$
and $\hat{S}_{j}^{\alpha}=g_{j}\otimes\vec{S}_{j}^{\alpha}g_{j}$. The minimum of the variational energy function corresponds to the ground state energy of the system.
Using the normalization condition $\langle \psi|\psi\rangle=1$ , we can mapped variational parameter  to
$a_{j}=\cos(\theta_{j})e^{i\phi_{j}}$, $b_{j}=\sin(\theta_{j})e^{\acute{\phi_{j}}}$. Therefor one can obtain
$\hat{S}_{j}^{z}=\frac{1}{2}\cos(2\theta_{j})$, $\hat{S}_{j}^{+}=(\hat{S}_{j}^{-})^{\dagger}=\frac{1}{2}\sin(2\theta_{j})e^{i\phi_{j}}$
where $\phi_{j}=\varphi_{j}-\acute{\varphi_{j}}$ and by choosing $\vec{D}=D\hat{z}$, it is found that:
\begin{eqnarray}
E_{var}&=&\frac{1}{4}\sum_{j}\{J\cos(2\theta_{j})\cos(2\theta_{j+1})\nonumber\\
       &+&D\sin(2\theta_{j})\sin(2\theta_{j})\sin(\phi_{j}-\phi_{j+1})\}.
\label{var}
\end{eqnarray}
By minimizing above equation, the ground state energy $(E_{GS})$ in the ferromagnetic
case $J < 0 $ shall be obtained. It was shown that the ground state energy has the constant
value, $E_{GS}=-\frac{N|J|}{4}$ for $D<|J|$ and decreasing linearly with DM interaction  for
$D>|J|$ as $E_{GS}=-\frac{ND}{4}$.
Now, buy using the minimized variational parameters of $E_{var}$  we are able to
calculate tangle Eq.(\ref{tangle2}) and concurrence Eq.(\ref{concurrence}). For $J>D$
one can obtain
\begin{eqnarray}
g^{xx}_{j,j+1}&=&g^{zz}_{j,j+1}=g^{yx}_{j,j+1}=g^{xy}_{j,j+1}=0\nonumber\\
g^{yy}_{j,j+1}&=&-1/4,
\label{var}
\end{eqnarray}
and $M^{z}=0$ so we have $\tau_{1}=1$ and $C_{j,j+1}=1/2$. For $J<D$ again using the above
conditions, one can obtain
\begin{eqnarray}
g^{xx}_{j,j+1}&=&g^{yy}_{j,j+1}=g^{yx}_{j,j+1}=g^{xy}_{j,j+1}=0\nonumber\\
g^{zz}_{j,j+1}&=&1/4,
\label{var}
\end{eqnarray}
and $M^{z}=1/2$ which give $\tau_{1}=0$ and $C_{j,j+1}=0$.

%%%%%%%%%%%%%%%%%%%%%%%%%%%%%%%%%%%%%%%%%%%%%%%%%
%%%%%%%%%%   FIG-5   %%%%%%%%%%%%%%%%%%%%%%%%%%

\begin{figure}
  \begin{center}
   \includegraphics[height=80mm,width=100mm]{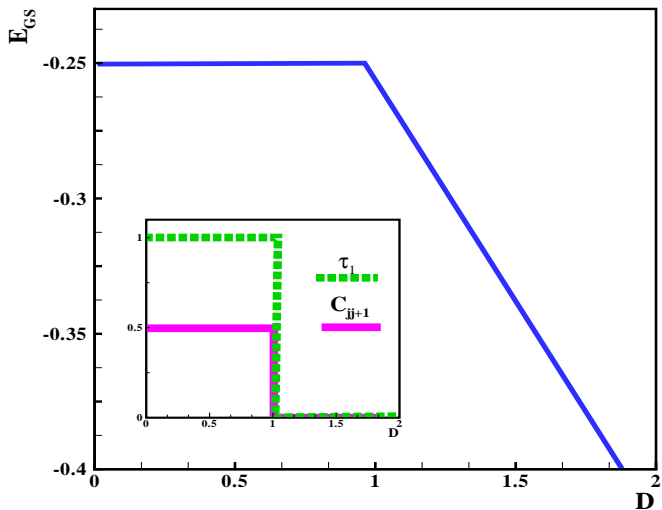}\\
     \caption{(Color online.) The ground state energy (normalized by N)
of ferromagnetic Ising chain with DM interaction using variational
matrix product state. Inset shows one-tangle and concurrence
aas function of the DM interaction using variational
matrix product state.}\label{variational}
\end{center}
\end{figure}

%########################################################################
%######################                        ##########################
%######################      Section IV       ##########################
%######################                        #########################
%#######################################################################

\section{THREE-QUBIT ENTANGLEMENT}\label{sec4 }

\begin{figure}
  \begin{center}
   \includegraphics[height=80mm,width=100mm]{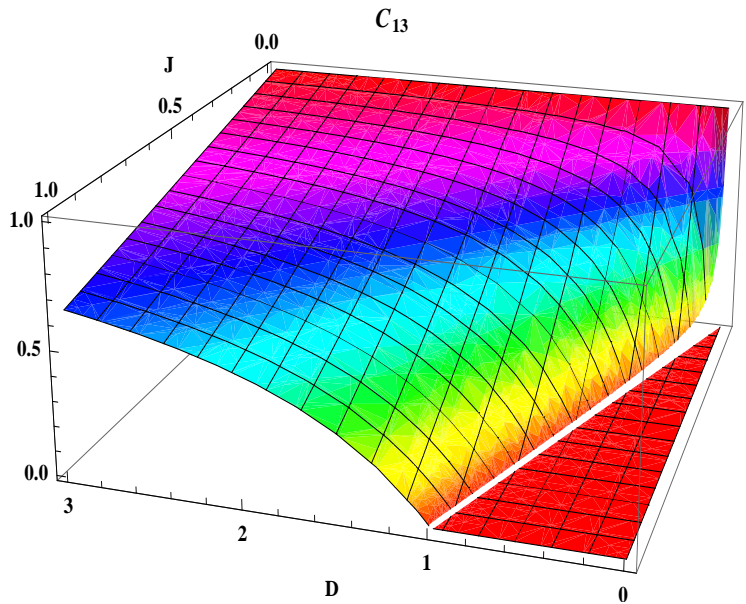}\\
     \caption{(Color online.) The ground state entanglement through symmetric way
,$C_{13}$, for antiferromagnetic case as function of DM and J exchange.}\label{C1-3T0}
\end{center}
\end{figure}

\begin{figure}
  \begin{center}
   \includegraphics[height=80mm,width=100mm]{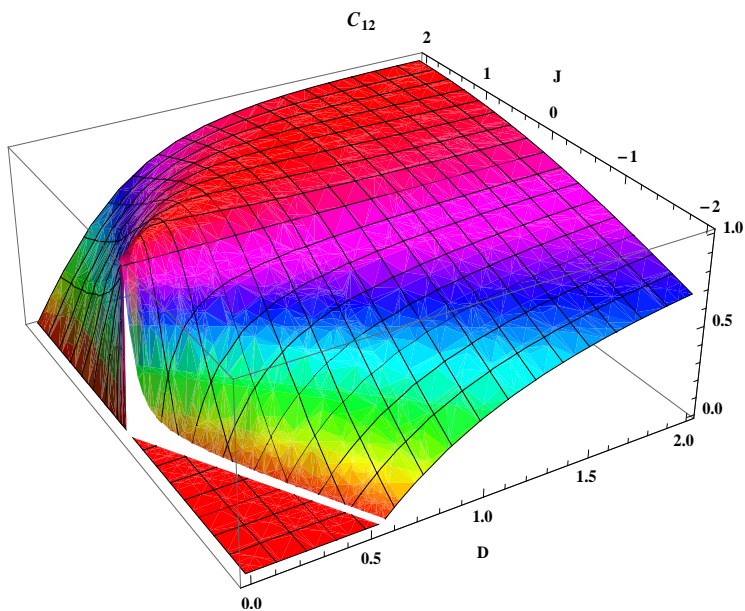}\\
     \caption{(Color online.) The ground state entanglement through non-symmetric way
,$C_{12}$, for both ferromagnetic an antiferromagnetic cases as function of DM and J exchange. }\label{C1-2T0}
\end{center}
\end{figure}

In this section we focus on the entanglement of formation of three-qubit in two inequivalent, symmetric and non-symmetric pairwise entanglement ways. By labeling the 3-qubits as $1, 2, 3$
sequentially. The symmetric reduced density matrix $\rho_{13}$ is defined as $\rho_{13}=tr_{2}(\rho)$,
where $\rho$ is the density matrix of 3-qubits. The non-symmetric reduced matrix is $\rho_{12}=tr_{3}(\rho)$.
Before present our results, we briefly review the definition of concurrence\cite{Wooters98,VCoffman00}.
Let $\rho_{ij}$ be density matrix of a pair of qubit $i$ and $j$. The concurrence corresponding to density
matrix is defined as
\begin{equation}
C_{ij}=max\{\lambda_{1}-\lambda_{2}-\lambda_{3}-\lambda_{4},0\},
\label{Cij}
\end{equation}
where the quantities $\lambda_{1}\geq\lambda_{2}\geq\lambda_{3}\geq\lambda_{4}$ are square roots of the eigenvalue of the operator
\begin{equation}
\varrho_{ij}=\rho_{ij}(\sigma_{y}\otimes\sigma_{y})\rho^{\ast}_{ij}(\sigma_{y}\otimes\sigma_{y}).
\label{roij}
\end{equation}
The concurrence $C_{ij}=0$ corresponds to an non-entangled state and  $C_{ij}=1$ corresponds to a maximally
entanglement sate. A straightforward calculation gives the eigenstates and the eigenvalues of 3-qubit of Eq.(\ref{ISING}).
The square roots of the operator $\varrho_{ij}$ for symmetric and nonsymmetric cases are presented in sequence.
In the symmetric case $\varrho_{13}$
\begin{eqnarray}
\lambda_{1}=0,\quad
\lambda_{2}=\lambda_{3}=\left( \frac{\alpha^{2}}{\alpha^{2}+2} \right)^{2},\quad\lambda_{4}=4\left( \frac{2}{\alpha^{2}+2} \right)^{2}
\end{eqnarray}
and for the non-symmetric case $\varrho_{12}$
\begin{eqnarray}
\lambda_{1}=\lambda_{2}=\left(\frac{1}{\alpha^{2}+2} \right)^{2},\quad\lambda_{3}=\left(\frac{(\alpha+1)^{2}}{\alpha^{2}+2} \right)^{2},\quad\lambda_{4}=\left(\frac{(\alpha-1)^{2}}{\alpha^{2}+2} \right)^{2}
\end{eqnarray}
where $\alpha=(J+\sqrt{8D^{2}+J^{2}})/2D$. In order to determine the existence of entanglement,
 we have considered ferromagnetic and antiferromagnetic
cases. Fig.\ref{C1-3T0} and Fig.\ref{C1-2T0} show symmetric, $C_{13}$,
and non-symmetric, $C_{12}$, concurrences. In the symmetric way, Fig.~\ref{C1-3T0}, it is
clear that in the absence of DM interaction the antiferromagnetic case is entanglement, in contrast to
ferromagnetic case which is fully unentangled. In antiferromagnetic case, the $C_{13}$ is zero up to $D_{c}=|J|$ and
starts to increase by increasing DM and reaches a saturation value. Our calculation shows a competition
between Ising  exchange $J$ and the DM interaction which entanglement starts to decreasing by increasing Ising exchange.
In the symmetric way, ferromagnetic Ising chain with added DM interaction does not
show any entanglement and by increasing DM interaction nothing will not happen.

%%%%%%%%%%%%%%%%%%%%%%%%%%%%%%%%%%%%%%%%%%%%%%%%%
%%%%%%%%%%   FIG-6   %%%%%%%%%%%%%%%%%%%%%%%%%%

%%%%%%%%%%%%%%%%%%%%%%%%%%%%%%%%%%%%%%%%%%%%%%%%%
%%%%%%%%%%   FIG-7   %%%%%%%%%%%%%%%%%%%%%%%%%%

In the non-symmetric way, Fig.~\ref{C1-2T0}, ferromagnetic Ising chain with added DM interaction does not
show any entanglement up to $D_{c}=|J|$. But for $D_{c}>|J|$ entanglement starts to increase by increasing DM interaction
and shows a competitive behavior between DM and J exchange which entanglement shows decreasing by increasing $|J|$.
 In antiferromagnetic chain, through non-symmetric case, opposite to ferromagnetic case, is fully entangled. Entanglement starts
 to increasing from zero as soon as turn on DM and reaches its saturation value around $D_{c}=|J|$. It should be mentioned that
 in antiferromagnetic case, increasing $J$ exchange can enhance the amount of entanglement.

\begin{figure}
  \begin{center}
   \includegraphics[height=95mm,width=110mm]{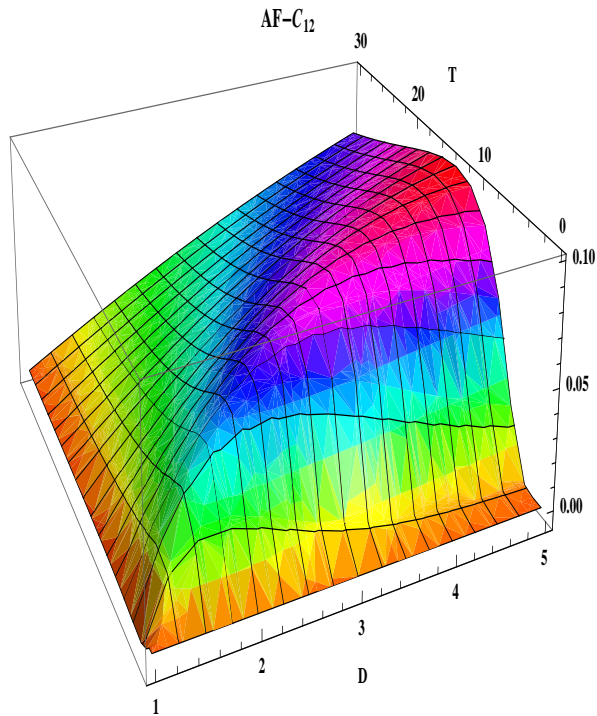}\\
     \caption{(Color online.) The thermal entanglement through non-symmetric way
,$C_{12}$, for  antiferromagnetic case as function of DM and temperature.}\label{thermal-C-AF}
\end{center}
\end{figure}

\begin{figure}
  \begin{center}
   \includegraphics[height=90mm,width=110mm]{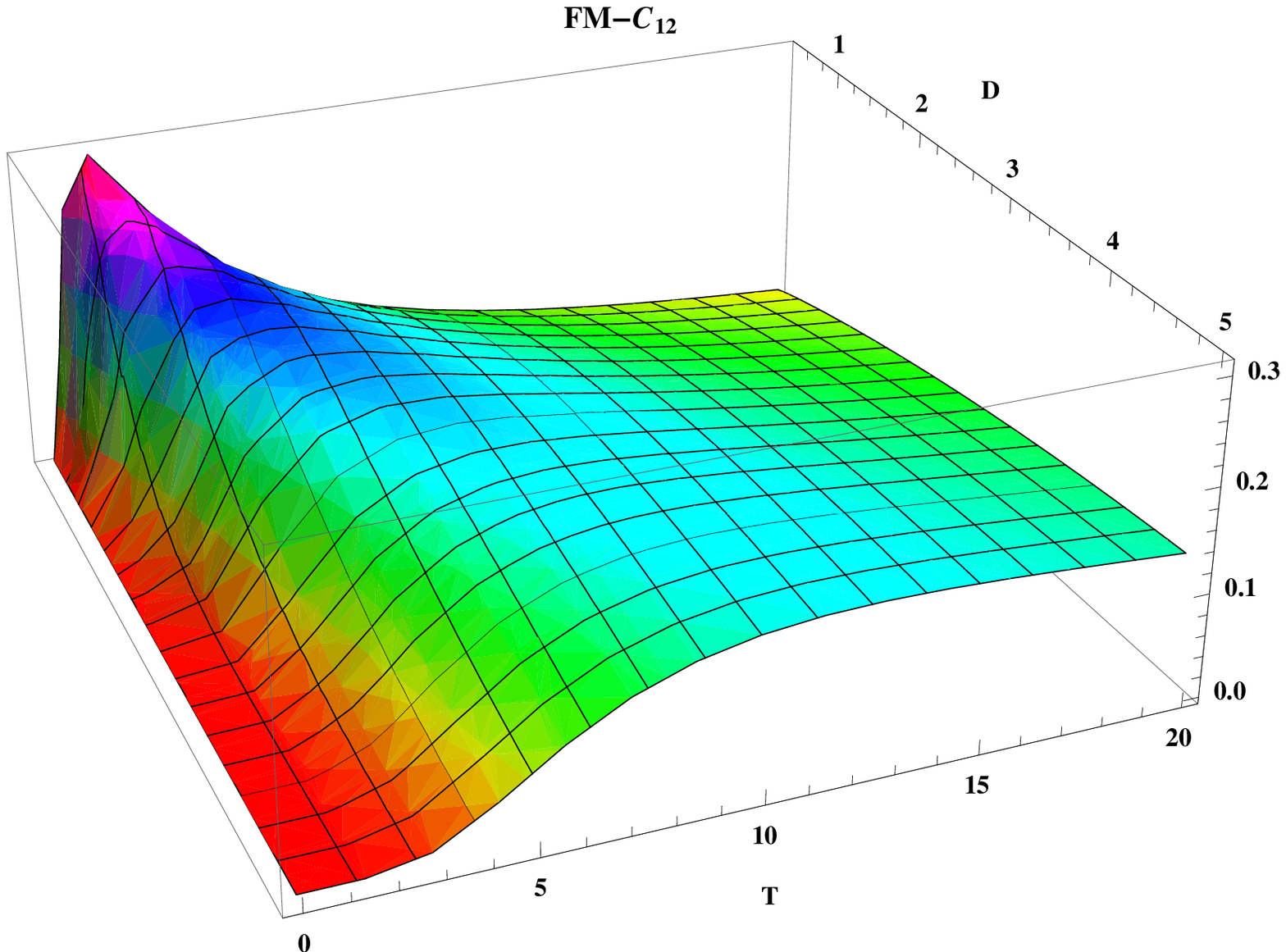}\\
    \includegraphics[height=90mm,width=110mm]{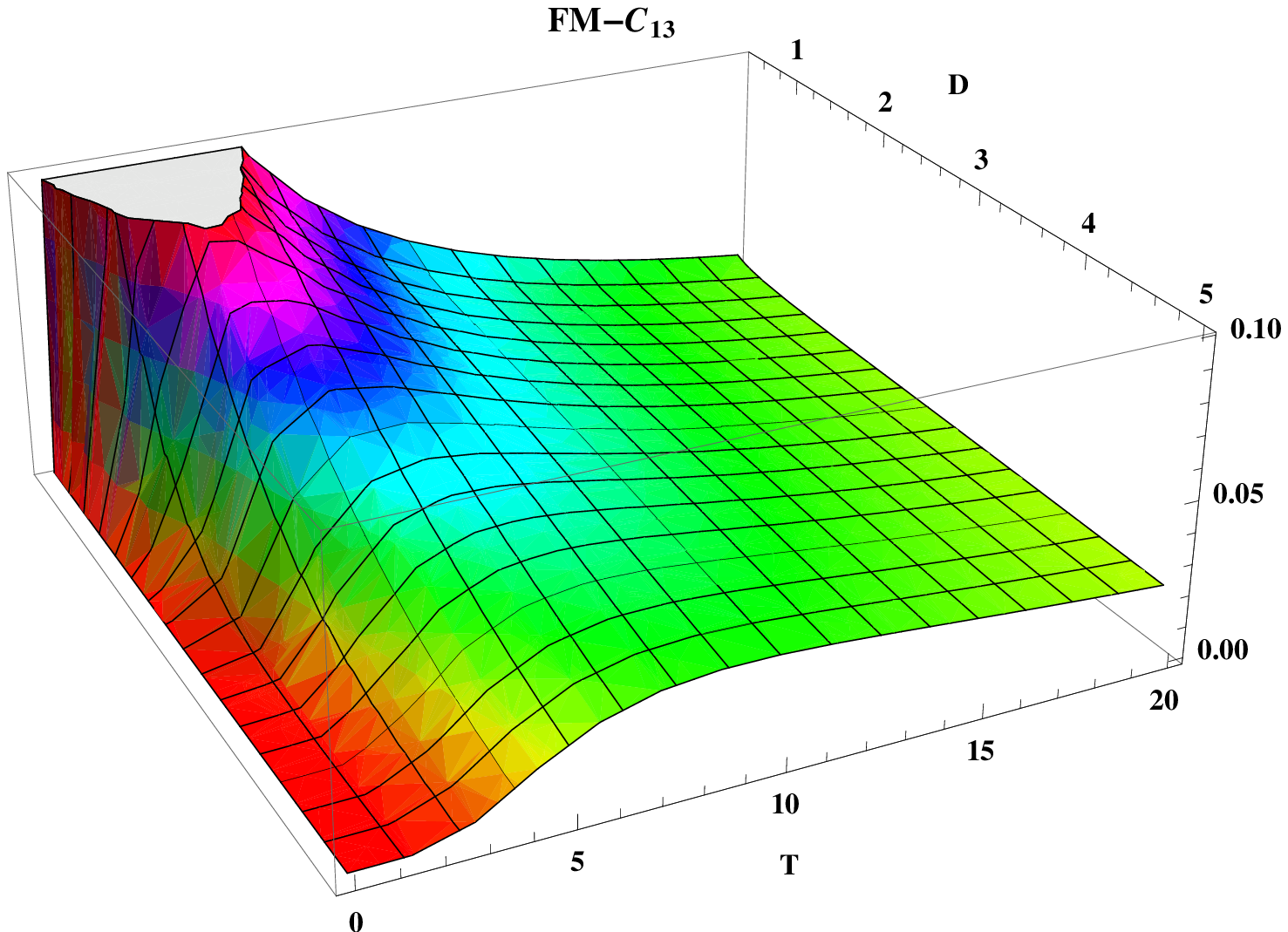}\\
     \caption{The thermal entanglement through non-symmetric way(top)
,$C_{12}$, and symmetric way(bottom) ,$C_{13}$,for ferromagnetic case as function of
DM and temperature. }\label{thermal-C-2-AF}
\end{center}
\end{figure}

The concept of thermal entanglement was introduced and studied within one-dimensional
isotropic Heisenberg model\cite{Arnesen00}. Here we study this kind of entanglement
within three-qubit Ising chain with added DM interaction through symmetric and non-symmetric way.
The state of the system at thermal equilibrium is $\rho(T)=\exp(-H/kT)/\Omega$, where $\Omega=Tr[\exp(-H/kT)]$ is
the partition function and $k$ is the Boltezmann's constant. As $\rho(T)$ represents thermal state,
the entanglement in the state is called thermal entanglement\cite{Arnesen00}.
The square roots of the operator $\varrho_{ij}(T)$ for symmetric and nonsymmetric cases are presented in sequence.
For the non-symmetric case $\varrho_{13}(T)$
\begin{eqnarray}
\lambda_{1}=\lambda_{2}=X_{12}^{2},\quad\lambda_{3}=\left( Y_{12}+Z_{12} \right)^{2},\quad
\lambda_{4}=\left( Y_{12}-Z_{12} \right)^{2}
\end{eqnarray}
where
\begin{eqnarray}
X_{12}&=&\frac{1}{\Omega}\left[\frac{1}{2}+\frac{e^{-\beta \varepsilon_{3}}}{a^{2}+2}+e^{-\beta \varepsilon_{5}}+\frac{e^{-\beta \varepsilon_{7}}}{b^{2}+2} \right],\nonumber\\
Y_{12}&=&\frac{1}{\Omega}\left[\frac{1}{2}+\frac{(a^{2}+1)e^{-\beta \varepsilon_{3}}}{a^{2}+2}+\frac{(b^{2}+1)e^{-\beta \varepsilon_{7}}}{b^{2}+2} \right],\nonumber\\
Z_{12}&=&\frac{1}{\Omega}\left[\frac{2ae^{-\beta \varepsilon_{3}}}{a^{2}+2}+\frac{2be^{-\beta \varepsilon_{7}}}{b^{2}+2} \right],
\end{eqnarray}
and for the non-symmetric case $\varrho_{12}(T)$
\begin{eqnarray}
\lambda_{1}=\lambda_{2}=X_{13}^{2},\quad\lambda_{3}=\left( Y_{13}+Z_{13} \right)^{2},\quad
\lambda_{4}=\left( Y_{13}-Z_{13} \right)^{2}
\end{eqnarray}
where
\begin{eqnarray}
X_{13}&=&\frac{1}{\Omega}\left[\frac{a^{2}e^{-\beta \varepsilon_{3}}}{a^{2}+2}+e^{-\beta \varepsilon_{5}}+\frac{b^{2}e^{-\beta \varepsilon_{7}}}{b^{2}+2} \right],\nonumber\\
Y_{13}&=&\frac{1}{\Omega}\left[1+\frac{(2e^{-\beta \varepsilon_{3}}}{a^{2}+2}+\frac{2e^{-\beta \varepsilon_{7}}}{b^{2}+2} \right],\nonumber\\
Z_{13}&=&\frac{1}{\Omega}\left[1-\frac{(2e^{-\beta \varepsilon_{3}}}{a^{2}+2}-\frac{2e^{-\beta \varepsilon_{7}}}{b^{2}+2} \right],
\end{eqnarray}
and $\Omega=4e^{-\beta J}[\cosh\beta J+\cosh\beta \sqrt{8D^{2}+J^{2}}]$, $a=J-\sqrt{8D^{2}+J^{2}}/2D$, $b=J+\sqrt{8D^{2}+J^{2}}/2D$.
In Fig.\ref{thermal-C-AF} and Fig.\ref{thermal-C-2-AF} we give two plots of the thermal concurrence of
antiferromagnetic and ferromagnetic Ising chains as functions of temperature
and DM interaction respectively. In antiferromagnetic case, Fig.\ref{thermal-C-AF}, model
dose not shows any entanglement through symmetric way and model lives in fully unentangled phase. In contrast to
symmetric way, non-symmetric way shows entanglement. As it can be seen from Fig.\ref{thermal-C-AF},
there is a region which surprisingly temperature can enhance entanglement and this region will become  wide spread for
higher DM. For higher temperature DM can not overcome temperature and entanglement gradually goes to zero by
increasing temperature.

%%%%%%%%%%%%%%%%%%%%%%%%%%%%%%%%%%%%%%%%%%%%%%%%%
%%%%%%%%%%   FIG-8   %%%%%%%%%%%%%%%%%%%%%%%%%%

%%%%%%%%%%%%%%%%%%%%%%%%%%%%%%%%%%%%%%%%%%%%%%%%%
%%%%%%%%%%   FIG-9   %%%%%%%%%%%%%%%%%%%%%%%%%%

In ferromagnetic case, Fig.\ref{thermal-C-2-AF}, both symmetric and non-symmetric
way have qualitatively the same behavior. Here, same as antiferromagnetic case, temperature can enhance entanglement.
Thermal and quantum fluctuations have a tight competition to drive
system in favorable regime, but in low temperature, both temperature and DM interaction has effective influence on the
degree of entanglement of the system and both of them  could be used to increase the entanglement of the spin system.\\
Either Fig.\ref{C1-3T0} and Fig.\ref{C1-2T0} are plotted for $D>D_{c}=1$, because we could not find any prominent behavior
for $D<D_{c}=1$. Actually both symmetric and non-symmetric entanglement are zero in the $D<D_{c}=1$ region.

%%%%%%%%%%%%%%%%

%%%%%%%%%%%%%%%%

 \section{CONCLUSION}\label{sec5 }

To summarize, we have investigated the effect  of a Dzyaloshinskii-Moriya (DM)
interaction on the ground state phase diagram of the one-dimensional (1D) Ising spin-1/2 model using
the variational matrix product state and numerical Lanczos methods from
entanglement point of view.
Our results show that there is a critical point in either ferromagnetic and antiferromagnetic
cases. In the ferromagnetic Ising chain this critical point was predicted
by variational matrix approach  exactly at $D_{c}=|J|$, which by using
numerical study we have confirmed it.  We have also  used the minimum of the
entanglement ratio $R\equiv\tau_{2}/\tau_{1}<1$,to check the presence of this quantum critical point.
For both ferromagnetic and antiferromagnetic
cases our numerical study gave the minimum of $R\equiv\tau_{2}/\tau_{1}$ at $D_c=|J|$,. We also used
generalized multipartite entanglement tools to check how entanglement is share in our model and check their ability to
detect critical point. For antiferromagnetic case we have calculated global $E_{gl}$ and generalized multipartite
entanglement $G(2,n)$ and either of them show that a quantum phase transition took place at $D_{c}=|J|$ .

We have calculated ground state entanglement of symmetric, $C_{13}$, and non-symmetric, $C_{12}$, concurrences for
 ferromagnetic and antiferromagnetic cases. In the symmetric way, in the absence of DM interaction
 both antiferromagnetic and ferromagnetic cases are  fully unentangled.
 By increasing DM, in antiferromagnetic case, the $C_{13}$ is zero up to critical
 points $D_{c}=|J|$. After the critical point entanglement starts to increase until its saturation.
 For ferromagnetic case, in symmetric way $C_{13}$ is always zero and
 does not show any entanglement.

 In the non-symmetric way, in the absence of DM interaction, $C_{12}$ is equal zero for
 either antiferromagnetic and ferromagnetic cases.  By turning DM nothing, in the ferromagnetic case,
 will not happen up to critical point $D_{c}$ and then after that   $C_{12}$ starts to increase by increasing
 DM interaction. In contrast to ferromagnetic case,
 in the antiferromagnetic case $C_{12}$ starts to increasing immediately after turning DM and
 reaches its saturation value around $D_{c}=|J|$.

We have also studied thermal entanglement of symmetric, $C_{13}(T)$, and non-symmetric, $C_{12}(T)$ for both
ferromagnetic and antiferromagnetic cases. Our calculations show that for either symmetric and non-symmetric cases
thermal(unentanglement favorable) and quantum(entanglement favorable) fluctuations have competition to drive system
to their favorable regime and surprisingly thermal decoherence can enhance entanglement in some part of low temperature region.

% %%%%%%%%%%%%%%%%%%%%%%%%%%%%%%%%%%%%%%%%%%%%%%%%%%%%%%%%%%%%%%%%%%%%%%%%%%%%%%%%%%%%%%%%%%
%\section{ACKNOWLEDGMENTS}
%It is our pleasure to thank ..., for very useful comments and
%interesting discussions.

%-----------------------------------------------------------------------------

\end{document}